\documentclass[10 pt, conference]{IEEEtran}
\ifCLASSINFOpdf
\else
\fi
\usepackage{amsmath,amsfonts}
\usepackage{epsf}
\usepackage{epsfig}
\usepackage{graphics}
\usepackage{graphicx}

\begin{document}

\title{Separation of Signals Consisting of Amplitude and Instantaneous Frequency RRC Pulses Using SNR Uniform Training}

\author{\IEEEauthorblockN{Mohammad Bari and Milo\v s Doroslova\v cki}}
\author{\IEEEauthorblockN{Mohammad Bari and Milo\v s Doroslova\v cki}\\
\IEEEauthorblockA{Department of Electrical and Computer Engineering\\
The George Washington University\\
Washington, DC, USA\\
\{mustafa,doroslov\}@gwu.edu}}

\maketitle

\begin{abstract}
This work presents sample mean and sample variance based features that distinguish continuous phase FSK from QAM and PSK modulations. Root raised cosine pulses are used for signal generation. Support vector machines are employed for signals’ separation. They are trained for only one value of SNR and used to classify the signals from a wide range of SNR. A priori information about carrier amplitude, carrier phase, carrier offset, roll-off factor and initial symbol phase is relaxed. Effectiveness of the method is tested by observing the joint effects of AWGN, carrier offset, lack of symbol and sampling synchronization, and fast fading.

\end{abstract}


%
\IEEEpeerreviewmaketitle

\section{Introduction}
\label{int}

Designing a classification algorithm to solve the Digital Modulation Classification (DMC) problem is a challenging undertaking. Three main issues should be considered. First, the signal of interest propagates through a channel which can affect the signal by distorting it with noise, Doppler shift and multipath fading. Second, one or more of the main parameters shaping the signal of interest can be unknown. Third, the complexity of the classification algorithm can be
high. This work concentrates on the simple solution to the preprocessing part in which continuous phase frequency shift keying (CPFSK) modulation is separated from linear modulations of phase shift keying (PSK) and quadrature amplitude modulation (QAM). All modulations use RRC shaped pulses. 
Wavelet-based feature is presented in \cite{ho00} for distinguishing frequency shift keying (FSK) from PSK having rectangular shaped pulses. Approximate entropy is exploited in \cite{Bari15ciss2}\nocite{pawar11}-\cite{Bari15asilomar2} to distinguish within the class of CPFSK having non-rectangular instantaneous frequency pulses. Neural network classifier is used in \cite{pawar11} and it is trained on 500 realizations. High training requirement coupled with the high number of samples per realization (8000) make it challenging for practical implementation, especially in real-time. In contrast, our method works well when FSK carriers are inseparable in the frequency domain, and when the number of samples taken in a symbol period is low, which ends up in processing fewer number of samples in a realization. This in turn makes real-time implementation less challenging. Furthermore, we use support vector machines (SVM) based classifier. In general, classifiers require the estimate of signal to noise ratio (SNR) to train itself. Therefore, classifiers are to be trained for every value of SNR. We, on the other hand, train SVM for only one value of SNR. The classification threshold computed for one value of SNR is then used to classify the signals having a large range of SNR. Low number of realizations (50) is used for training. To the best of our knowledge there is no published work on how to separate CPFSK signals from QAM and PSK modulations where all modulations use RRC shaped pulses, and where training is as short as presented in this work. Standard deviation based features presented in \cite{Bari15asilomar3} separate frequency modulated signals from linear modulations for rectangular shaped pulses.

Our choice of the features is motivated by the work presented in \cite{kay89}, where a simple frequency estimator is proposed using the phase of a product of consecutive complex samples. In our case, the goal is to distinguish between linear and non-linear digital modulations and instead of using the phase of a product of consecutive complex samples, we use the sample mean and sample variance of imaginary part of the product of consecutive complex samples. For rectangular pulses, the sample mean of the imaginary part of the product is discussed in specific DMC scenarios in \cite{Bari14asilomar}\nocite{Bari13ciss}-\cite{Bari13asilomar} where only binary FSK (BFSK) modulation is separated from linear modulations. These pulses are not used in majority of modern systems. Additionally, instead of SVM, threshold based classification is discussed in those papers. Since a lot of a priori information on signal parameters is needed to calculate thresholds, the approach is of less practical interest. In \cite{Bari15spl}\cite{Bari15cssp}, the exact knowledge of the SNR of the received signal is assumed. Reference \cite{pauluzzi00} surveys SNR estimation techniques. The knowledge of the value of SNR is relaxed by 3 dB in \cite{Bari15ciss1}.

In sections \ref{sig}, \ref{der} and \ref{sim} the baseband discrete-time signal model used in
this paper, the proposed features and simulation experiments are discussed, respectively.

\section{Signal Model}
\label{sig}

Consider a scenario where spectrum of a signal is observed and its center is estimated (not considered in this paper). Then the signal's spectrum is translated to be around a certain desired normalized frequency, say $0$ or $\pi/2$. The obtained signal should be classified as CPFSK or linearly modulated one. Let complex baseband continuous-time received signal be
\begin{equation}
s(t)=
x(t-t_0)e^{j(\Delta t+\theta_c)}\alpha(t)e^{j\psi(t)}+v(t)
\label{e1}
\end{equation}
where $x(t)$ is the transmitted-signal part of the received
signal, $t_0$ is the time delay, $\theta_c$ is the initial phase
uniformly distributed over $[0,~2\pi)$, $\Delta$ is the
carrier offset, $(\alpha(t),~\psi(t))$ are the
amplitude and phase of the multiplicative noise used here to
model fading. $\alpha(t)$ is assumed to be a Rayleigh random
variable, while $\psi(t)$ is independent of $\alpha(t)$ and
uniformly distributed over $[0,2\pi)$. $v(t)$ is assumed to be zero-mean complex noise and $t_0$ is time delay. We assume
that the bandwidth $B$ of the receiver's filter is in general
larger than the bandwidth of the transmitted signal and the
power spectral density of $v(t)$ is constant within the filter
bandwidth. In the case of QAM and PSK modulations the
transmitted-signal part of the received signal is

\begin{equation}
x(t)=
\sum\limits_{n=-\infty}^{+\infty}a_ne^{j\theta_n}p(t-nT)
\label{e2}
\end{equation}
where $(a_n,~\theta_n)$ are the amplitude and phase of the transmitted symbol, $p(t)$ is the pulse shape function and $T$ is the symbol period. In the case of FSK modulation

\begin{equation}
x(t)=
e^{j\int\limits_{0}^{t}\omega(\rho)d\rho}
\label{e3}
\end{equation}
where $\omega(\rho)=\sum_{n=-\infty}^{+\infty}b_nq(\rho-nT)$ is the instantaneous frequency of the modulated signal
where $q(\rho)$ defines instantaneous frequency pulse shape and $b_n\in\{-1,+1\}$ for BFSK.
Note that (\ref{e3}) models continuous phase FSK, which is of higher practical interest than non-continuous phase FSK used in \cite{Bari13acceptedAsilomar}. 

Now, we assume that we are sampling with period $T_s=1/(2B)$. Symbol period, $T$, is given by $T=N_sT_s+\varepsilon T_s$ where $N_s$ is the number of sample periods per symbol period and $\varepsilon\in[0,1)$.
The complex baseband discrete-time received signal
is
\begin{equation}
s[k]\hspace{-1mm}=\hspace{-1mm}
x(kT_s-t_0)e^{j(\Delta' k+\theta_c)}\alpha[k]e^{j\psi[k]}\hspace{-1mm}+v[k]\hspace{-1mm}=\hspace{-1mm}s(t)\vert_{t=kT_s}\label{e13}
\end{equation}
where $\Delta'=\Delta T_s$, $\alpha[k]=\alpha(kT_s)$, $\psi[k]=\psi(kT_s)$ and $v[k]$ is complex circular AWGN having zero mean. Note that $t_0=k_0T_s+\varepsilon_0T_s$ where $k_0$ is the integer part and $\varepsilon_0\in[0,1)$ is the
fractional part of the time delay $t_0$ measured in sampling
periods as time units.

\section{Proposed Features}
\label{der}

Motivated by \cite{kay89} we consider
\begin{equation}
\ w[k]=s[k]s^*[k-1]
\label{e14}
\end{equation}
where * is the complex conjugate operator. From now on, let us
assume that there is no
fading, i.e., $\alpha[k]=1,~\psi[k]=0$. Now, applying $w[k]$ on noiseless QAM
and PSK signals yields
\begin{equation}
\begin{split}
\hspace{-2mm}
w[k]\vert_{v[k]\equiv 0} &=\hspace{-2mm}
\sum\limits_{n=-\infty}^{+\infty}a_n^2e^{j\Delta'}P_n[k]P_n^-[k]+
\\&~~~\hspace{-5mm}
\sum\limits_{n=-\infty}^{+\infty}\sum\limits_{m=-\infty}^{+\infty}a_na_me^{j[\theta_n-\theta_m+\Delta']}P_n[k]P_m^-[k]
\end{split}
\label{e15}
\end{equation}
where $P_n[k]P_m^-[k]=p(kT_s-t_0-nT)p(kT_s-T_s-t_0-mT).$

\hspace{-5mm}Similarly, applying $w[k]$ on noiseless FSK signals yields 
\begin{equation}
\begin{split}
w[k]\vert_{v[k]\equiv 0}&=
e^{j[\int\limits_{kT_s-T_s-t_0}^{kT_s-t_0}\omega'(\rho)\frac{d\rho}{T_s}+\Delta']}
\\&\hspace{-12mm}
=e^{j\sum\limits_{n}b_n\int\limits_{kT_s-T_s-t_0}^{kT_s-t_0}q(\rho-nT)d\rho+j\Delta'}\hspace{-4mm}=e^{j\sum\limits_{n}b_nQ_n[k]+j\Delta'}
\end{split}
\label{e16}
\end{equation}
where $Q_n[k]=\int_{kT_s-T_s-t_0}^{kT_s-t_0}q(\rho-nT)d\rho.$

Let us consider the means of imaginary part of $w[k]$ in (\ref{e15}) and (\ref{e16}) for linear and BFSK modulations, respectively. For equiprobable constellation points of each modulation, mean of imaginary part of $w[k]$ for 16-QAM, BPSK, 4-PSK and 8-PSK signals is given by 
\begin{equation}
\begin{split}
\hspace{-4mm}E[\mbox{Im}(w[k]\vert_{v[k]\equiv 0})]=&E[\mbox{Im}(w[k])]
\\&\hspace{-4mm}
=\mbox{sin}(\Delta')\sum\limits_{n=-\infty}^{+\infty}P_n[k]P_n^-[k]E[a_n^2]
\label{e17}
\end{split}
\end{equation}
where 16-QAM has constellation points $a_ne^{j\theta_n}\in\{k/\sqrt{10}+jl/\sqrt{10};~k,~l=-3,-1,+1,+3\}$, BPSK's phases $\theta_n\in\{0,~\pi\}$, 4-PSK's phases $\theta_n\in\{(2n+1)\pi/4;~n=0,1,2,3\}$, and 8-PSK's phases $\theta_n\in\{(2n+1)\pi/8;~n=0,1,...,7\}$. Constellation points for 16-QAM, BPSK, 4-PSK and 8-PSK are chosen such that the average power is unity. Mean of imaginary part of $w[k]$ in (\ref{e16}) for BFSK signal is 
\begin{equation}
E[\mbox{Im}(w[k]\vert_{v[k]\equiv 0})]\hspace{-1mm}=\hspace{-1mm}E[\mbox{Im}(w[k])]\hspace{-1mm}=\hspace{-1mm}\sin(\Delta')\hspace{-3mm}\prod_{m=-\infty}^{\infty}\hspace{-3mm}\cos(Q_m[k]).
\label{e18}
\end{equation}

Next, let us consider the variances of imaginary part of $w[k]$ in (\ref{e15}) for QAM and PSK modulations, and in (\ref{e16}) for BFSK modulation. The variance for BPSK is 
\begin{equation}
\begin{split}
\hspace{-2mm}\mbox{VAR}[\mbox{Im}(w[k]\vert_{v[k]\equiv 0})]&\hspace{-1mm}=\hspace{-1mm}
\sin^2(\Delta')\big[(\hspace{-2mm}\sum\limits_{m=-\infty}^{+\infty}\hspace{-3mm}P_m[k]P_m^-[k])^2+
\\&
\hspace{-18mm}\hspace{-3mm}\sum\limits_{m=-\infty}^{+\infty}\hspace{-3mm}P_m^2[k]\hspace{-3mm}\sum\limits_{m=-\infty}^{+\infty}\hspace{-3mm}(P_m^-[k])^2\hspace{-1mm}-\hspace{-1mm}2\hspace{-3mm}\sum\limits_{m=-\infty}^{+\infty}\hspace{-3mm}(P_m[k]P_m^-[k])^2\big].
\end{split}
\label{e18b}
\end{equation}
For QAM, 4-PSK and 8-PSK modulations, whose signal constellations are invariant to $\pi/2$ rotation, the variance is 
\begin{equation}
\begin{split}
\mbox{VAR}[\mbox{Im}(w[k]\vert_{v[k]\equiv 0})]\hspace{-1mm}&=\hspace{-1mm}
\sin^2(\Delta')\bigg[(E[a_0^4]\hspace{-1mm}-\hspace{-1mm}2E^2[a_0^2])\hspace{-3mm}
\\&\hspace{-31mm}
\sum\limits_{m=-\infty}^{+\infty}\hspace{-3mm}(P_m[k]P_m^-[k])^2
\hspace{-1mm}+\hspace{-1mm}E^2[a_0^2](\hspace{-2mm}\sum\limits_{m=-\infty}^{+\infty}\hspace{-3mm}P_m[k]P_m^-[k])^2\hspace{-0.5mm}\bigg]\hspace{-1.5mm}
+\hspace{-1.1mm}\frac{1}{2}E^2[a_0^2]\times
\\&
\hspace{-29mm}
\bigg[\sum\limits_{m=-\infty}^{+\infty}P_m^2[k]\sum\limits_{m=-\infty}^{+\infty}(P_m^-[k])^2-(\sum\limits_{m=-\infty}^{+\infty}P_m[k]P_m^-[k])^2\bigg].
\end{split}
\label{e18a}
\end{equation}
The variance of $\mbox{Im}(w[k])$ in (\ref{e16}) for BFSK modulation is
\begin{equation}
\begin{split}
\hspace{-2mm}\mbox{VAR}[\mbox{Im}(w[k]\vert_{v[k]\equiv 0})]\hspace{-1mm}&=\hspace{-1mm}\frac{1}{2}\hspace{-1mm}-\hspace{-1mm}\frac{1}{2}\hspace{-2mm}\prod\limits_{m=-\infty}^{+\infty}\hspace{-2mm}\cos(2Q_m[k])+
\\&
\hspace{-25mm}\sin^2(\Delta')\big(\hspace{-2mm}\prod\limits_{m=-\infty}^{+\infty}\hspace{-2mm}\cos(2Q_m[k])\hspace{-1mm}-\hspace{-2mm}\prod\limits_{m=-\infty}^{+\infty}\hspace{-2mm}\cos^2(Q_m[k])\big).
\end{split}
\label{e18c}
\end{equation}

For rectangular pulses $p(t)$ and $q(t)$ with support on $[0,T)$, amplitudes 1 and $\beta'/T_s$, respectively, unit average power signal constellations, and $\epsilon=\epsilon_0=0$, $E[\mbox{Im}(w[k])]$ in (\ref{e17}) and (\ref{e18}) for linear and BFSK modulations, respectively, are
\begin{equation}
E[\mbox{Im}(w[k]\vert_{v[k]\equiv 0})]=(1-\frac{1}{N_s})\sin(\Delta')
\label{e17a}
\end{equation}
\begin{equation}
\hspace{-2mm}E[\mbox{Im}(w[k]\vert_{v[k]\equiv 0})]=\cos(\beta')\sin(\Delta').
\label{e17b}
\end{equation}
For $\Delta'=\pi/2$, it is $1-1/N_s$ for linear modulations and $\cos(\beta')$ for BFSK modulation. Therefore, the mean of $\mbox{Im}(w[k])$ at $\Delta'=\pi/2$ can be used to distinguish between the linear and the BFSK modulations. Similarly, the $\mbox{VAR}[\mbox{Im}(w[k]\vert_{v[k]\equiv 0})]$ in (\ref{e18b}) for BPSK, in (\ref{e18a}) for 4-PSK, 8-PSK and 16-QAM, and in (\ref{e18c}) for BFSK modulations, respectively, are
\begin{equation}
\begin{split}
\hspace{-3mm}\mbox{VAR}[\mbox{Im}(w[k]\vert_{v[k]\equiv 0})]&\hspace{-1mm}=\hspace{-1mm}\frac{1}{N_s}\sin^2(\Delta')\hspace{-1mm}=\hspace{-1mm}
\left\{
\begin{array}{ll}
\hspace{-2mm}0,~~~\Delta'\hspace{-1mm}=\hspace{-1mm}0
\\
\hspace{-2mm}\frac{1}{N_s},\Delta'\hspace{-1mm}=\hspace{-1mm}\pi/2
\end{array}
\right.
\end{split}
\label{e19}
\end{equation}
\begin{eqnarray}
\lefteqn{\hspace{-13mm}\mbox{VAR}[\mbox{Im}(w[k]\vert_{v[k]\equiv 0})]\hspace{-1mm}=\hspace{-1mm}(1\hspace{-1mm}-\frac{1}{N_s})\sin^2(\Delta')(E[a_0^4]\hspace{-1mm}-\hspace{-1mm}1)\hspace{-1mm}+\hspace{-1mm}\frac{1}{2N_s}}\nonumber
\\& &
\hspace{-11mm}=\hspace{-1mm}\left\{
\begin{array}{l}
\hspace{-2mm}\frac{1}{2N_s},\Delta'=0
\\
\hspace{-2mm}\frac{1}{2N_s}+(1-\frac{1}{N_s})(E[a_0^4]-1),\Delta'=\pi/2
\end{array}
\right.
\label{e20}
\end{eqnarray}
\begin{equation}
\begin{split}\hspace{-15mm}
\mbox{VAR}[\mbox{Im}(w[k]\vert_{v[k]\equiv 0})]&=
(1\hspace{-1mm}-\hspace{-1mm}\sin^2(\Delta'))\sin^2(\beta')
\\&
=\left\{
\begin{array}{ll}
\hspace{-2mm}\sin^2(\beta'),&\Delta'\hspace{-1mm}=\hspace{-1mm}0
\\
\hspace{-2mm}0,&\Delta'\hspace{-1mm}=\hspace{-1mm}\pi/2
\end{array}
\right.
\end{split}
\label{e21}
\end{equation}
where $E[a_0^4]$ is 1.32 for 16-QAM and it is 1 for 4-PSK and 8-PSK modulations. For $\beta'$ such that $\sin^2(\beta')>1/(2N_s)$ at $\Delta'=0$, the variance of the BFSK modulations is larger than the variance of linear modulations. On the other hand, the variance of the BFSK modulation is smaller than the variance of linear modulations at $\Delta'=\pi/2$. Therefore, the sample variances of $\mbox{Im}(w[k])$ at $\Delta'=0$ and $\Delta'=\pi/2$ can be jointly used together with the sample mean of $\mbox{Im}(w[k])$ at $\Delta'=\pi/2$ to distinguish the linear modulations from the BFSK modulation in the case of rectangular pulses. Furthermore, it can be shown that the features can be used to separate $\it{L}$-ary-FSK from linear modulations.  

\section{Simulations and Discussion} 
\label{sim}

In this section, simulation experiments are presented illustrating the classification performance of the proposed features for RRC shaped pulses. The performance for the proposed features is also compared to the performance of a classifier using the wavelet-based feature for
distinguishing PSK from FSK in \cite{ho00}. We modify this approach by incorporating SVM in classification \cite{burges98}. The number of training data used to construct the classifiers for the two classes is 50 realizations of each modulation. For wavelet-based approach, SVM is trained for every value of SNR. For the proposed features, SVM is trained for only 10 dB. The corresponding SVM threshold is then used for classifying signals having several values of SNR. The performance of the proposed features classified by SVM that is trained for only one value of SNR is then compared to the performance of the proposed features classified by SVM that is trained for every value of SNR.

Error performance of the methods is compared in the presence of additive white Gaussian noise (AWGN), unknown carrier offset, asynchronous sampling and fast fading. For both the methods there are 70000 modulated signals, 10000 signals for each modulation (16-QAM, BPSK, 4-PSK, 8-PSK, BFSK, 4-FSK and 8-FSK). There are 600 symbols in one realization. Discrete-time signals are obtained by taking only 2 samples per symbol period ($N_s$=2). Carrier offset, $\Delta'$, and roll-off of RRC pulses are fixed for a realization and they are uniformly distributed in $[\gamma'-\pi/20$, $\gamma'+\pi/20]$, where $\gamma'\in\{0,~\pi/2\}$ and in $\{k/10;~k=1,~2,~3,....,~10\}$, respectively. The symbol period and time delay are non-integer multiples of sampling period. This results in asynchronicity between sampling instants and symbol period. $\varepsilon$ and $\varepsilon_0$ are both uniformly distributed in [0,1) and remain unchanged for a particular realization, that is, $\varepsilon\ne0$, $\varepsilon_0\ne0$. $k_0$ is uniformly distributed in $\{0, 1, 2, ..., \lceil N_s+\epsilon\rceil-1\}$. For the fast fading, $\alpha[k]e^{i\psi[k]}$ is such that its autocorrelation magnitude remains over 1/2 of the fading average power for 9 samples, which is around four times more than $N_s=2$ in our simulations. $\alpha[k]$ is chosen to have unit mean square value, that is, $E[\alpha^2[k]]=1$. For optimal results of wavelet/SVM approach, the signals' spectra are placed around normalized frequency $\pi/2$. The scale for the continuous wavelet transform is 2 and the Haar wavelet is used as the mother wavelet.  The length of median filter is 2 for wavelet-based method.  

\begin{figure}[t]
\centerline{{\epsfysize=7.25cm\epsfbox{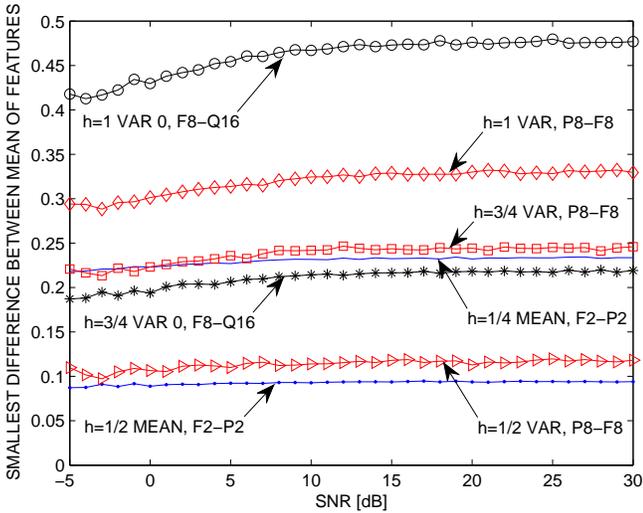}}}
\caption{Smallest difference in sample means and sample variances of $\mbox{Im}(w[k])$ between FSK and linear modulations.}
\label{figvar}
\end{figure}

First, the empirical evidence is discussed to prove the usefulness of the features for RRC pulses in the joint presence of AWGN, carrier offset, asynchronicity and fast fading. Figure \ref{figvar} shows the difference in the closest sample means and sample variances of $\mbox{Im}(w[k])$ between FSK and linear modulations. ``$h$=x VAR (MEAN), P8-F8'' corresponds to the sample variance (mean) of $\mbox{Im}(w[k])$ for 8-PSK minus that of 8-FSK with $h$=x for carrier frequency centered at $\pi/2$. ``$h$=x VAR 0, F8-Q16'' corresponds to the sample variance of $\mbox{Im}(w[k])$ for 8-FSK with $h$=x minus that of 16-QAM for carrier frequency centered at 0. Pm, Qm and Fm represent $\it{L}$-ary PSK, QAM and FSK, respectively. It can be seen that the separation, which translates into feature's effectiveness, in sample mean (variance) based feature increases as $h$ decreases (increases). For $h\le1/2$, the sample variance based features for $\gamma'=0$ and $\pi/2$ are not too helpful. However, in this range of $h$, the mean based feature is effective. Similarly, sample variance based features help in classification when sample mean based feature fails ($h>1/2$). 

\begin{figure}[t]
\centerline{{\epsfysize=7.25cm\epsfbox{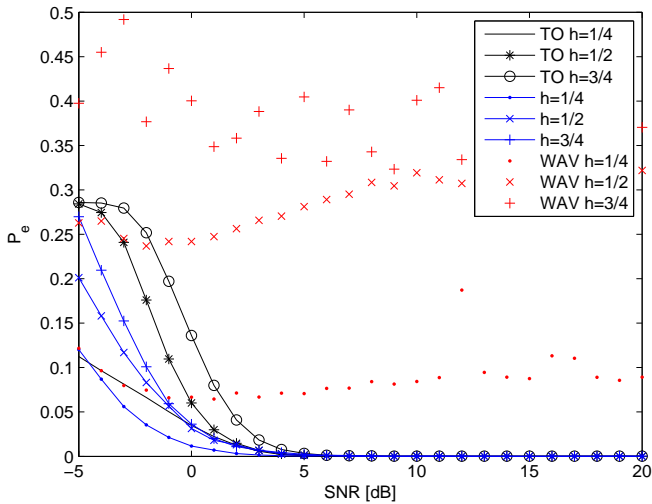}}}
\caption{Probability of misclassification ($P_e$). ``TO $h$=x'' represents the proposed features for $h$=x where SVM is trained only once at 10 dB, ``(WAV) $h$=x'' represents the (wavelet-based feature) proposed features for $h$=x where SVM is trained for every value of SNR.}
\label{fig_hFixed}
\end{figure}

The performance of the two methods is shown in Figure \ref{fig_hFixed} for $h\in\{1/4,1/2,3/4\}$. Here fast fading is present along with AWGN, carrier offset and asynchronicity. Note that before further processing the received signal is normalized by multiplying it with $\sqrt{(1+\sigma^2)/P_r}$ where $\sigma^2$ is the noise power and $P_r$ is the received signal's average power. Estimating noise power is a classical problem and it is outside the scope of this paper. A survey of SNR estimation techniques is carried out in \cite{pauluzzi00}. It is worth mentioning that from the spectrum of  BFSK signal the existence of two carrier frequencies is not observable even for $h=3/4$. It is clear from the figure that the performance of proposed features trained for only one value of SNR is comparable to that where SVM is trained for every value of SNR, espacially for SNR$>$4 dB. Based on the SNR of the signals, SVM creates a classification threshold. For separable case, SVM tends to maximize the threshold distance from the closest samples of the two classes in feature space. The distance increases as the SNR increases and hence the signals can be classified easily for higher values of SNR and vice versa. The simulation results show that the classification threshold that SVM generates at 10 dB is good enough to classify the signals for the adjacent and higher values of SNR. Though, SVM are trained for every value of SNR, wavelet/SVM simply fails to separate signals for small $N_s$. Note that the probability of error for both the methods decrease by increasing $N$ or $N_s$.

\section{Conclusion}
\label{con}
In this paper we propose simple and yet robust features to distinguish CPFSK modulations from QAM and PSK modulations that use RRC pulses in the joint presence of AWGN, carrier offset, lack of synchronization and fast fading. The features are based on the sample mean and sample variance of imaginary part of the product of two consecutive signal values. Probability of error for the features with SVM trained for only one value of SNR is comparable to the case where SVM is trained for every value of SNR. Moreover, no a priori information is required about carrier amplitude, carrier phase, carrier offset, symbol rate, pulse shape and initial symbol phase (timing offset).

Simulation results showed that for quite low oversampling the proposed classifier performs well while wavelet-based classifier simply fails to classify. In order to have better error performance, the number of samples per symbol and/or symbols have to be increased. Furthermore, the number of samples needed for reliable classification by the proposed features is less than that of wavelet feature.

%
%
%
%
%
\bibliographystyle{IEEEtran}
\bibliography{bibAug5_2015}
\end{document}